\documentclass[showpacs,oneside,twocolumn,prl,amsmath,amssymb,fleqn]{revtex4-1}
\usepackage{cases}
\usepackage{amsmath}
\usepackage{amssymb}
\usepackage{amsfonts}
\usepackage{amssymb}
\usepackage{dcolumn}
\usepackage{bm}
\usepackage{bbm}
\usepackage{graphicx}
\usepackage{xcolor}
\usepackage{array}
\usepackage{subfigure}

\begin{document}
\title{Experimental protection and revival of quantum correlation in open solid systems}
\author{Xing Rong}
\author{Fangzhou Jin}
\author{Zixiang Wang}
\author{Jianpei Geng}
\author{Chenyong Ju}
\author{Ya Wang}
\author{Runmin Zhang}
\author{Changkui Duan}
\author{Minjun Shi}\altaffiliation{Corresponding author: shmj@ustc.edu.cn}
\author{Jiangfeng Du}\altaffiliation{Corresponding author: djf@ustc.edu.cn}
\affiliation{Hefei National Laboratory for Physics Sciences at the Microscale and Department of Modern Physics, University of Science
and Technology of China, Hefei, 230026, China}

\begin{abstract}
Quantum correlation quantified by quantum discord has been demonstrated experimentally as important physical resources in quantum computation and communication for some cases even without the presence of entanglement. However, since the interaction between the quantum system and the noisy environment is inevitable, it is essential to protect quantum correlation from lost in the environment and to characterize its dynamical behavior in the real open systems. Here we showed experimentally in the solid-state P:Si system the existence of a stable interval for the quantum correlation in the beginning until a critical time $t_c \approx  166$ ns of the transition from classical to quantum decoherence.  To protect the quantum correlation, we achieved the extension of the critical time by $50$ times to $8~\mu$s by applying a two-flip dynamical decoupling (DD) pulse sequence. Moreover, we observed the phenomenon of the revival of quantum correlation, as well as classical correlation. The experimental observation of a non-decay interval for quantum correlation and the great extension of it in an important solid-state system with genuine noise makes the use quantum discord as physical resources in quantum information processing more practicable.

\end{abstract}

\maketitle

Quantum correlations that can be shared only among quantum systems play essential roles in fundamental physics and future technologies \cite{Horodecki2009}. Although quantum entanglement has been widely investigated, it cannot describe the nonclassicality of the correlations in separable states\cite{Bennett1999}, while quantum discord can \cite{Ollivier2002,Henderson2001}. A great deal of theoretical investigations \cite{Datta2008,Cavalcanti2011,Cornelio2011,Werlang2010,Streltsov2012,Chuan2012,Rossatto2011,Perinotti2012,Madhok2012} on quantum discord have been carried out recently to show the properties and importance of quantum discord. Moreover, quantum discord has been demonstrated experimentally as important resources in deterministic quantum computation with one pure qubit (DQC1) \cite{Lanyon2008PRL} and remote state preparation even without the presence of entanglement \cite{Dakic2012natphys}. Furthermore, an operational method to use quantum discord (without the presence of entanglement) as a physical resource has been demonstrated in Ref. \cite{Gu2012natphys}. Since all quantum systems inevitably interact with the environment, the evolution of the quantum discord in a noisy environment is certainly of great interests. It is well-known that in a dissipative environment entanglement decays to zero in a finite time, which is called entanglement sudden death (ESD) \cite{Yu2009science}. In contrast, quantum discord has been claimed to have no sudden death \cite{Werlang2009PRA}, but has peculiar sudden change in its decay rates \cite{Maziero2009PRA}. It has also been shown that quantum discord is robust in its initial period of decoherence but suffers a sudden change phenomenon \cite{Mazzola2010PRL}. Such a behavior has been recently experimentally investigated in optics \cite{XJS2010natcomun} and liquid NMR systems \cite{Auccaise2011PRL}, but the dynamics of quantum discord in solid-state systems with a real noisy environment remains elusive.

Meanwhile, it is essential to preserve the quantum correlation in a fragile quantum system, thus it is vital to overcome decoherence effect induced by the environment. One of the strategies is the dynamical decoupling (DD) technique \cite{Viola1999PRL,Uhrig2007PRL}, which uses stroboscopic spin flips to reduce the average coupling to the environment to effectively zero. DD is a particularly promising strategy for combating decoherence \cite{Du2009nature,Wang2011PRL}, since it can be naturally integrated with other desired functionalities, such as quantum gates.

Herein, we investigate the dynamics of both the classical and quantum correlations in solids. We find that in the initial interval, the quantum correlation is stable while the classical correlation decays. After this time interval, the quantum correlation decays, while the classical correlation remains. Furthermore, we manage to extend the time point of the sudden transition from the classical decoherence to the quantum decoherence is delayed from about $t_c \approx 166~$ns to about $8~\mu$s by DD. This time point can be further extended to about $12~\mu$s by a multiple two-flip DD pulse sequence. Additionally, we have observed the revival of both the classical and the quantum correlations, which may have some impact on quantum information processing.

For a biparticle state $\rho_{AB}$, the total correlation is quantified by mutual information $I(\rho_{AB})=S(\rho_A)+S(\rho_B)-S(\rho_{AB})$, where $S(\rho)=-\text{Tr}[\rho \text{Log}_2\rho]$ is the von Neumann entropy. $\rho_A$ and $\rho_B$ are the reduced-density matrices of $\rho_{AB}$. When performing a positive-operator-valued measure \{$\Pi_k  ^B$\} on particle $B$, the maximal accessible information about particle $A$ gives the classical correlation $C_B(\rho_{AB})=\max_{ \{\Pi_k  ^B\}}[S(\rho_A)-\sum_k p_k S(\rho^{k}_A)]$,
where $p_k=\text{Tr}[(1 \otimes \Pi_k ^B) \rho_{AB} (\mathbbm1 \otimes \Pi_k  ^B)^\dagger]$, $\rho^{k}_A=\text{Tr}_B[(\mathbbm1 \otimes \Pi_k  ^B) \rho_{AB} (\mathbbm1 \otimes \Pi_k  ^B)^\dagger]/p_k$.
Hence, the quantum part of the total correlation is defined as quantum discord $D_B(\rho_{AB})=I(\rho_{AB})-C_B(\rho_{AB})$. Note that quantum discord is asymmetric, i.e. generally $D_B(\rho_{AB})\neq D_A(\rho_{AB})$. Herein, for simplicity we consider an initial states of the form of the Bell-diagonal states,
\begin{equation}
\label{Bell-diagonal-states}
\rho_{AB} = \frac{1}{4}(\mathbbm{1} + \sum^3_{i = 1} c_i \sigma_i^A\sigma_i^B ),
\end{equation}
where $\sigma_i^{A(B)}$($i = 1, 2, 3$) are the Pauli matrices. For the density matrices of the Bell-diagonal states, Ref.~\cite{Luo2008PRA} and \cite{Shi2011NJP} provided the analytical expression and the geometric picture of the quantum discord, respectively.

In our experiment, we have chosen phosphorous donors in silicon (P:Si) \cite{Tyryshkin2003PRB,Wang2011PRL,Rong2012PRB} material with P concentration about $1 \times 10^{16}~$cm$^{-3}$ for the study. Silicon is a particularly attractive material for hosting spin qubit for its low-orbit coupling and low natural abundance of nuclear-spin-bearing isotope \cite{Morton2011Nature}. The experiment was performed at temperature 8 K. P:Si consists of an electron spin, $S=1/2\ (g=1.9987)$, coupled isotropically to the nuclear spin, $I=1/2$, of $^{31}$P, with a hyperfine coupling constant $A=117~$ MHz. The energy diagram of this system is plotted in Fig.\ref{fig1}a, where the four-level system can be manipulated by resonant the microwave (MW1, MW2) and radio-frequency (RF1, RF2) radiation. The dynamics of an open quantum system $\rho_{AB}$ coupling to the environment in solids can be described by longitudinal and transverse relaxations of time constants $T_1$  and $T_2$, respectively.
 For the electron spin, the transverse relaxation time is $T_{2e} = 120~\mu$s, the longitudinal (i.e., electron population) relaxation time is $T_{1e} = 5.6~$ms and the dephasing time $T_{2e}^*\approx 0.2 ~\mu$s. For the nuclear spin, the dephasing time is determined as $T_{2n}^* = 24~\mu$s using the nuclear spin free induction decay (FID) experiment.

The longitudinal relaxation time $T_1$ of both the electron and the nuclear spins are much larger than the transverse relaxation time $T_2$ and so can be neglected in our experiments. The decay of the off-diagonal elements depends on the time scale of dephasing time $T_2^\ast$ of the electron and the nuclear spin. Since the electron dephasing time is almost two orders of magnitude smaller than the nuclear dephasing time, the decay of the secondary diagonal elements is dominated by $T_{2e}^\ast$ (see Supplementary Information for details). For the phosphorous donors in silicon with natural abundance of $^{29}$Si, the hyperfine fields of the $^{29}$Si nuclei cause random static shifts of the individual electron-spin resonant frequencies which satisfy a Gaussian distribution \cite{Abe2010PRB}. Averaging over the Gaussian-distributed resonant frequencies, the dephasing of the electron spin can be derived with the form $\sim \exp[-(t/T_{2e}^\ast)^2]$. As a result, the time evolution of the total system is given by
\begin{equation}
  \label{Bell-diagonal-states-t}
\rho_{AB}(t)= \frac{1}{4}[\mathbbm{1} + c_1 (t) \sigma_x^A\sigma_x^B + c_2 (t) \sigma_y^A\sigma_y^B + c_3 (t) \sigma_z^A\sigma_z^B],
\end{equation}
where the coefficients $c_1 (t)=c_1 (0) \exp[-(t/T_{2e}^\ast)^2]$, $c_2 (t)=c_2 (0) \exp[-(t/T_{2e}^\ast)^2]$ and $c_3 (t)=c_3 (0)=c_3$.

The typical coefficients $c_i$ initialized by electron spin resonance (using $ \sim 10 \rm{GHz}$ excitation at a temperature of $8~\rm{K}$) are  $\sim 10^{-3}$, therefore we focus on a class of reasonable and widely used states for which $c_1 (0)=0, |c_3 (0)| < |c_2 (0)| \ll 1$. By following Ref.~\cite{Luo2008PRA}, we obtain the analytical expressions for the mutual information, the classical correlation and the quantum discord. Expanding $I[\rho(t)]$ and $C[\rho(t)]$ in the Taylor series of $c_2 (t)$ and $c_3$ and neglecting high-order terms, we obtain
\begin{equation}
  \label{mutual information}
 I[\rho(t)]=\frac{1}{2\ln2}~[c^2_3+c^2_2(t)],
\end{equation}
\begin{equation}
   \label{classical correlation}
C[\rho(t)]=\frac{1}{2\ln2}~c^2(t),
\end{equation}
where  $c(t)=\max\{c_2(t),c_3\}$. Hence, the quantum discord is calculated to be
\begin{equation}
 \label{quantum discord}
D[\rho(t)]=
   \begin{cases}
  \frac{1}{2\ln2}~c^2_3 & \text{if $t\leqslant t_c$},\\
  \frac{1}{2\ln2}~c^2_2 (t) & \text{if $t>t_c$}.
  \end{cases}
\end{equation}
Here $t_c= \sqrt {-\ln[c_3/c_2(0)]} T_{2e}^\ast$ is obtained by setting $c_2 (t_c)=c_3$. Consequently, in the noisy environment, the quantum discord is constant and the classical correlation decreases in the initial period $t\leqslant t_c$ , while for $t>t_c$, the classical correlation does not change in time and only the quantum discord is reduced sharply.

\begin{figure}
\centering
\includegraphics[width=0.9\columnwidth]{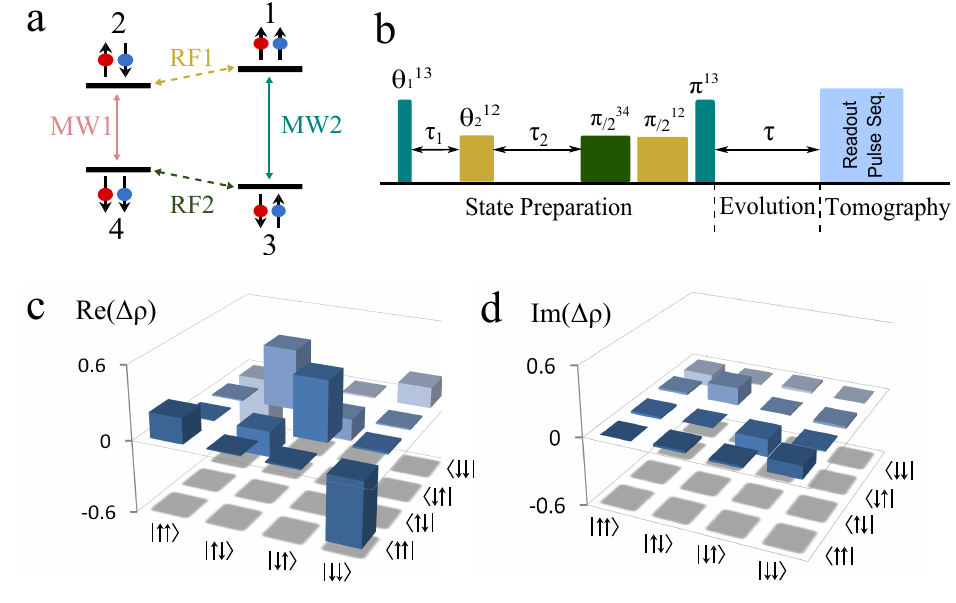}
   \caption{\label{fig1}{\bf Energy level diagram, experimental pulse sequence and reconstructed deviation density matrix.}
{\bf a,} Energy level diagram for the P:Si system. There are four Zeeman product states which are labeled by states $1- 4$, respectively. $\uparrow$ and $\downarrow$ stand for the $\pm 1/2$ states of electron and nuclear spins. EPR and NMR transitions are indicated by two-way arrows.
{\bf b,} Diagram of the experimental pulse sequence, which includes three steps: initial state preparation, relaxation delay and final state detection.
{\bf c, d, } The real and imaginary parts of the reconstructed deviation density matrix $\Delta \rho = \rho - \frac{1}{4}$, in unit of $\varepsilon$, respectively, where $\varepsilon = g \beta_e B_0/ {8 k_B T} = 7.35\times 10^{-3}$ (at temperature 8 K) is the ratio between the magnetic and thermal energies.
}
\end{figure}

\begin{figure}
\centering
\includegraphics[width=0.9\columnwidth]{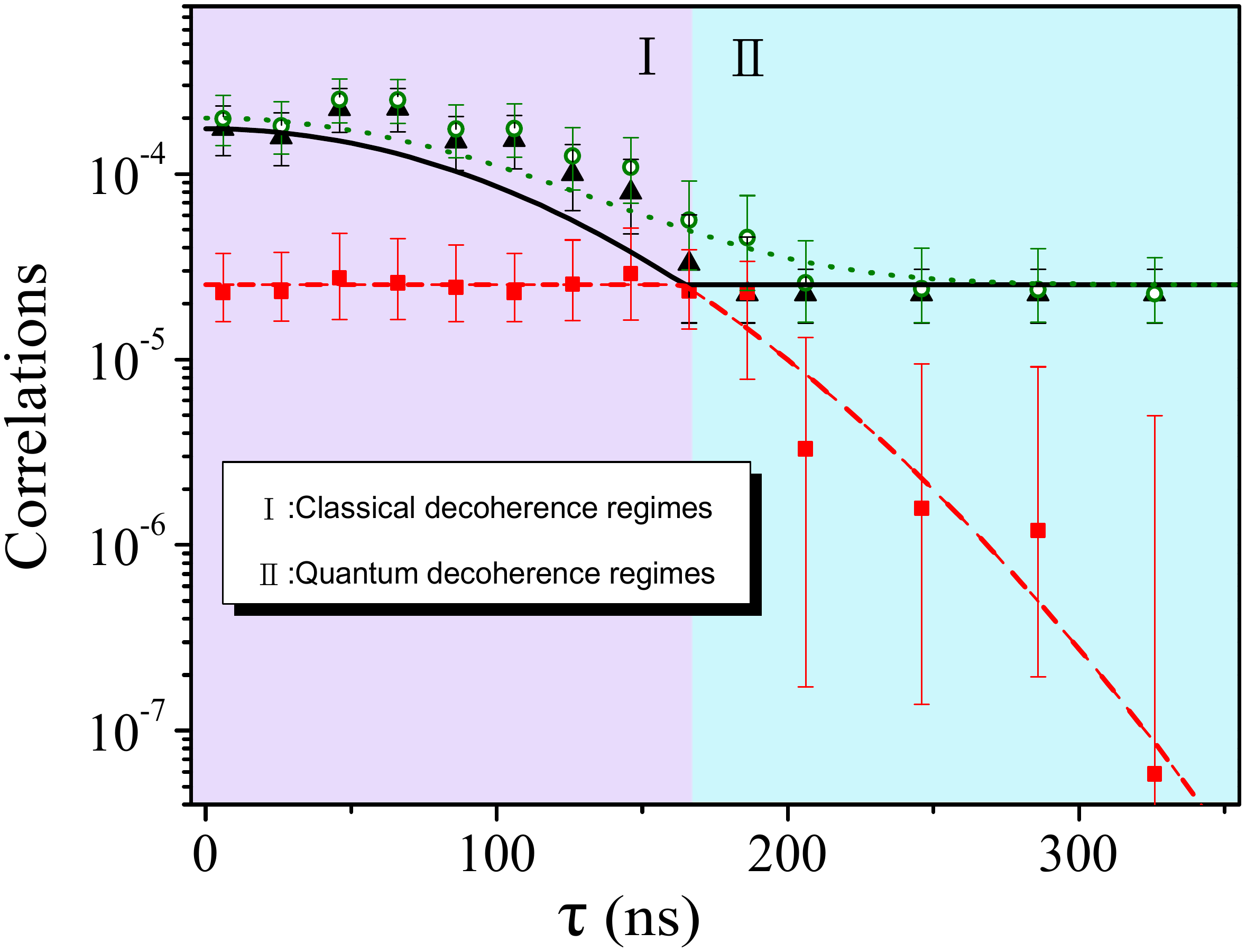}
    \caption{\label{fig2}{\bf Dynamics of classical and quantum Correlations.}
    The logarithmic-scale values of mutual information (darkgreen circle), quantum discord (red square) and classical correlation ( black triangle) are numerical computed with their original definitions (and so do the other experimental data in the article) (see Supplementary Information for details). The curves are the theoretical predication according to Eq. (\ref{quantum discord}). \uppercase\expandafter{\romannumeral1} and \uppercase\expandafter{\romannumeral2} stand for the classical decoherence and quantum decoherence regimes, respectively. Error bars are calculated (see Supplementary Information for details).
 }
\end{figure}

  In our experiment, the duration of the microwave and the radio-frequency $\pi$ pulses are $60$ ns and $10~\mu$s, which are determined using the EPR and NMR Rabi nutation experiments, respectively. By following Ref.\ \cite{Tyryshkin2003PRB}, the nuclear population relaxation time $T_{1n}$ is estimated to be 250 times of $T_{1e}$. So the waiting time between each experiment is set to 10 s. Fig.\ref{fig1}b shows the first pulse sequence applied in our experiment. It consists of three steps: the initial state preparation, the relaxation delay and the final state detection. Starting from the thermal equilibrium state, a MW2 pulse is used to flip the electron spin with an angle $\theta _1$ while the nuclear spin $I_z = 1/2$. This is followed by a waiting time of ($\tau_1 = 1~\mu$ s $\gg T_{2e}^*$), so that the off-diagonal elements of the density matrix decay off. After that the $\pi/2$ RF2 (RF1) pulse and the $\pi$ MW2 pulses are applied to generate the nonzero secondary element. Since the pulses $\pi/2$ RF1 and $\pi/2$ RF2 are applied sequentially, after generating the off-diagonal elements $\rho_{34}$ and $\rho_{43}$ by RF2, they will decay during the period of the next pulse RF1 with $T_{2n}^*$. To compensate for this effect, we add a $\theta _2$ RF1 between the first pulse MW2 and the duration $\tau_2$ ($\tau_2 = 200~\mu$s $\gg T_{2e}^*$), as shown in the Fig.\ref{fig1}b (see Supplementary Information for details). In our experiment, the actual values of $\theta _1$ and $\theta _2$ are chosen to be $\theta _1\thickapprox 0.70~\pi$ and $\theta _2\thickapprox 0.28~\pi$. Then the initial state is prepared as a Bell-diagonal state, which satisfies the conditions $c_1 (0)=0, |c_3 (0)| < |c_2 (0)| \ll 1$. We present the result of the tomography of the initial state in Fig.\ref{fig1}c,d, which represent the real and imaginary parts of the deviation density matrix in unit of $\varepsilon$. After the preparation step, the quantum system is left to evolve under the noisy environment. Then the resultant quantum states are reconstructed by state tomography \cite{Rong2012PRB,Vandersypen2004RMP}.

In our case, according to Eqs.\ (\ref{classical correlation}) and (\ref{quantum discord}), the dynamics of the quantum and classical correlations depend on the decay of secondary diagonal elements. The decay of $\rho_{23}$ and $\rho_{14}$ is shown in Supplementary Information with the fitted decay function of $\sim \exp[-(t/T_{\rm decay})^2]$, where $T_{\rm decay}=175$ ns. The values of mutual information (darkgreen circle), the quantum discord (red quadrate) and the classical correlation (black triangle) calculated from the state tomography results (see Supplementary information for details) are plotted in Fig.\ref{fig2}. The theoretical predications plotted in Fig.\ref{fig2} can successfully describe the experimental data and clearly show that the sudden transition from the classical to the quantum decoherence regime occurs at about 166 ns. In the initial period ($0 \leq t\lesssim 166$ ns), the quantum discord remains constant but the classical correlation decreases, while for $t\gtrsim166$ ns, the classical correlation does not change but the quantum correlation decreases dramatically.

\begin{figure}
\centering
\includegraphics[width=0.9\columnwidth]{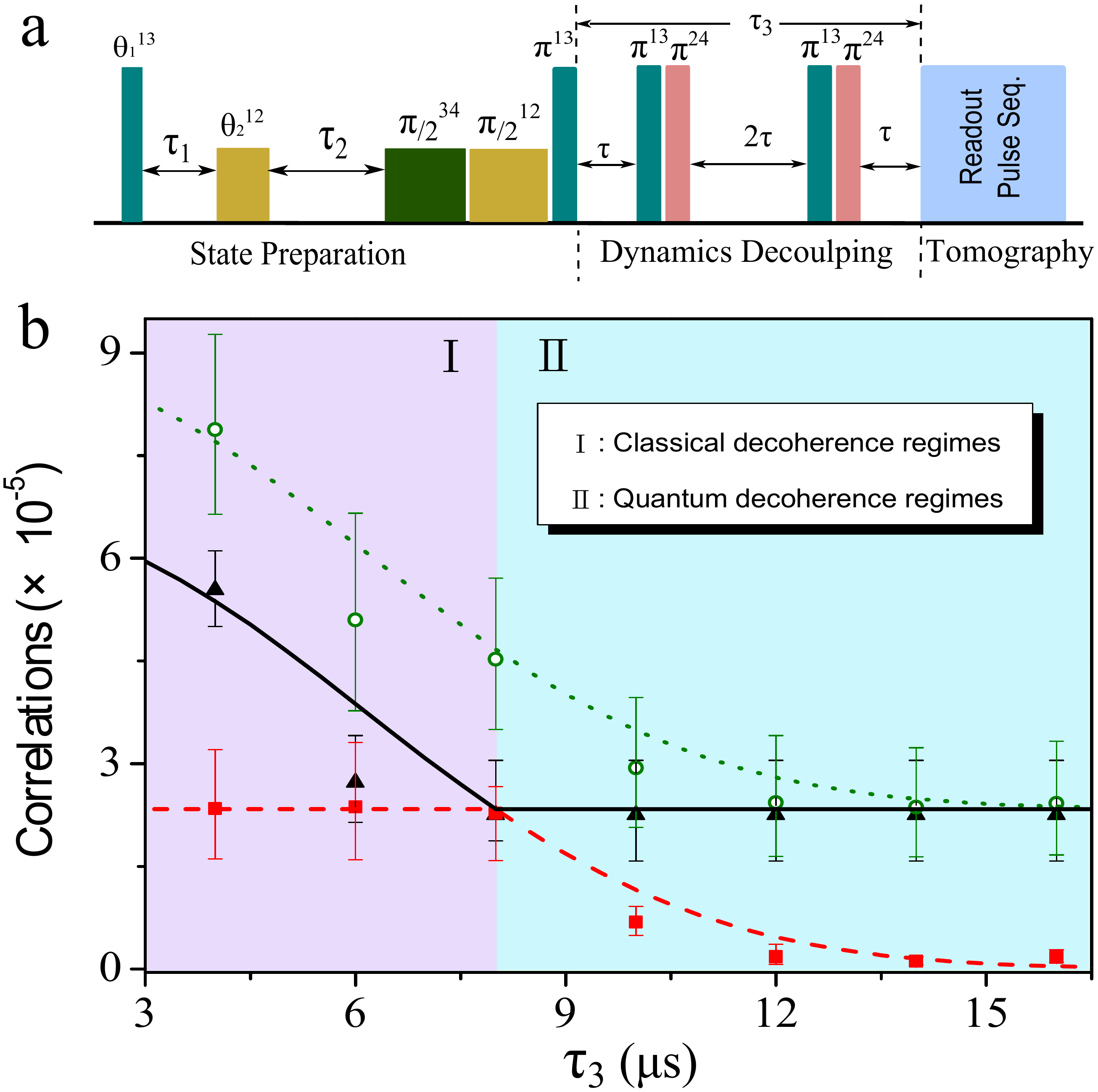}
    \caption{\label{fig3} {\bf Preservation of classical and quantum correlations by dynamics decoupling.}
    {\bf a,} Schematic illustration of the dynamics decoupling pulse sequence. After the initial state preparation, a set of two-flip DD sequence pulses for the electron spin are applied at $\tau$ and $3\tau$, and then the state tomography is performed at $4\tau$. {\bf b,} The plot of the results using the same notation as in Fig.2. The curves are drawn to follow the trend of the variation of correlations. \uppercase\expandafter{\romannumeral1} and \uppercase\expandafter{\romannumeral2} denote the classical and quantum decoherence regimes, respectively.}

\end{figure}

Dynamics decoupling is an effective method to preserve the coherence of a quantum system, and it has been tested by experiments in both single qubit \cite{Du2009nature} and two-qubit  \cite{Wang2011PRL} systems. Herein we use the two-flip DD sequence pulses for the electron spin to prolong both classical and quantum correlations. Fig.\ref{fig3}a shows the sequence pulses used in the dynamics decoupling experiment. After preparing the initial state, the two-flip DD sequence pulses for the electron spin are applied, and the finally state tomography is performed at $4\tau$. Fig.\ref{fig3}b shows that the decay of classical and quantum correlations become much slower, and the period before the sudden transition of correlations is prolonged via DD by about 50 times.

\begin{figure}
\centering
\includegraphics[width=0.9\columnwidth]{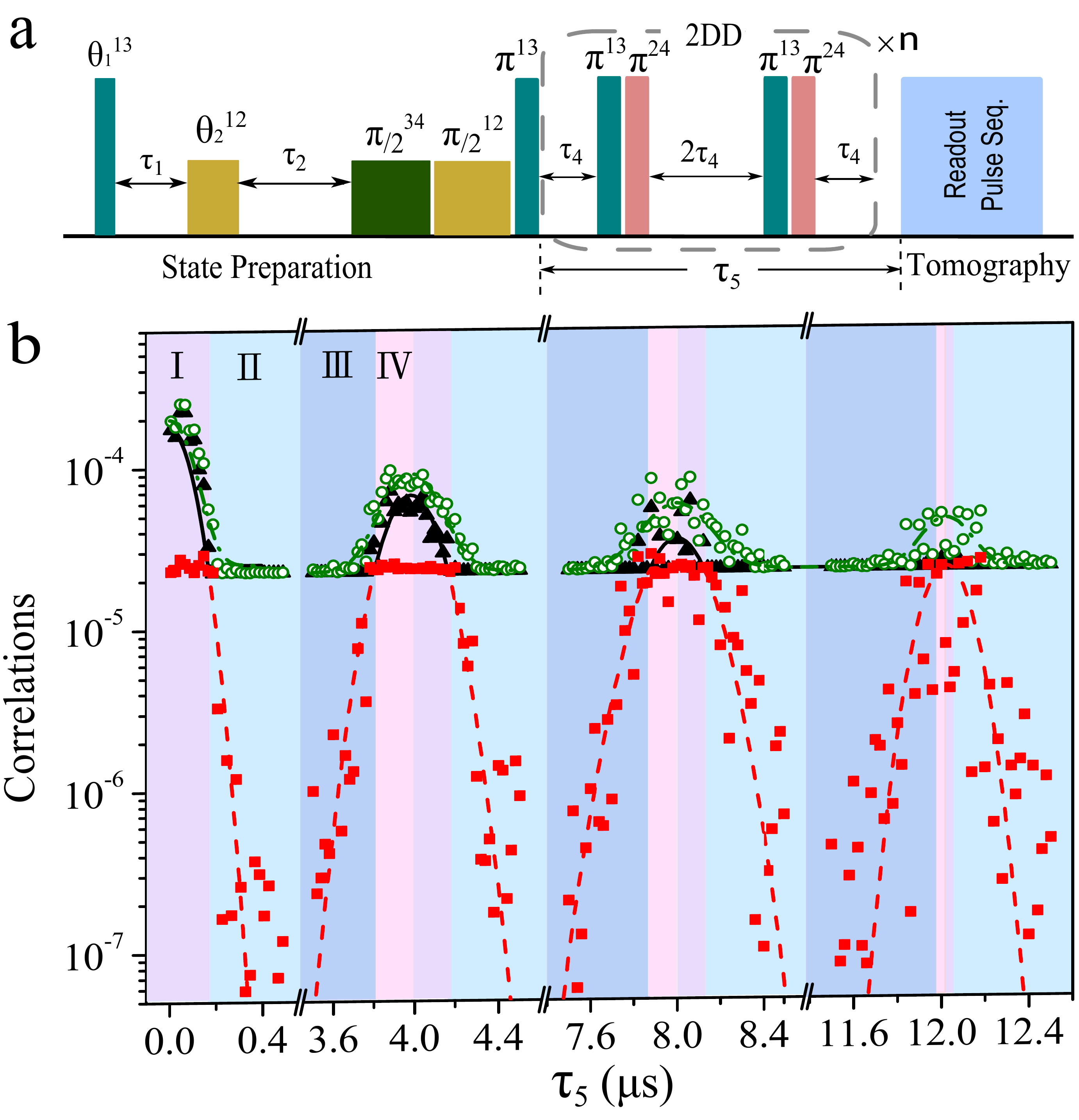}
    \caption{\label{fig4} {\bf Revival of classical and quantum correlations by dynamics decoupling pluses.}
      {\bf a,} Schematic illustration of the revival experiment pulse sequences.  After the initial state preparation, multiple two-flip DD pulses ($\tau_4=1~\mu$s) for the electron spin are applied. {\bf b,}  The plot of the results in logarithmic scale using the same notation as in Fig.2.  \uppercase\expandafter{\romannumeral1} and \uppercase\expandafter{\romannumeral2} stand for the classical decoherence and quantum decoherence regimes, respectively. \uppercase\expandafter{\romannumeral3} and \uppercase\expandafter{\romannumeral4} stand for the quantum revival and classical revival regimes, respectively.
 }
\end{figure}

After the dephasing, the classical and quantum correlations decay to almost zero, but do not disappear completely. When a two-flip DD sequence pulse for the electron spin is applied after a period $\tau_4$, as the Fig.\ref{fig4}a shows ($\tau_4=1~\mu$s), along with another same two-flip DD sequence pulse for the electron spin at time $3\tau_4$, there will be a revival of the classical and the quantum correlations at about the time $4\tau_4$. This process could be repeated several times, and it is observed that both the classical and the quantum correlations vanish and then revive. Fig.\ref{fig4}b shows the experimental observation, it is quite clear that the quantum discord (red square) decays from a constant value to nearly zero and then revive to a stable value at different stages by applying a proper two-flip DD pulse sequence, while the classical correlation (black triangle) decreases dramatically from an initial value to a constant value and has a similar revival. \uppercase\expandafter{\romannumeral1} and \uppercase\expandafter{\romannumeral2} stand for the classical decoherence and quantum decoherence regimes, respectively. The sudden transition from a quantum revival regimes (labeled by $\rm{\uppercase\expandafter{\romannumeral3}}$ ) to a classical revival regimes (labeled by $\rm{\uppercase\expandafter{\romannumeral4}}$) is obtained in Fig. \ref{fig4}b. Moreover, we can see that the time interval before the sudden transition of the correlation is prolonged to about $12~\mu$s by the sequence of the three two-flip DD pulses, longer than the time by applying a single two-flip DD showed in Fig.\ref{fig3}b.

We are aware of that there are other measures for quantum correlations \cite{Modi2012arXiv}, among which is the geometric measure of the quantum discord \cite{Dakic2010PRL}, which differs in general from the conventional definition of quantum discord \cite{Ollivier2002,Henderson2001}. This geometric measure of quantum discord has been shown \cite{Dakic2012natphys} to be related to the fidelity of quantum remote state preparation. For the two-qubit system studied here, the dynamical behavior of the geometric measure of quantum discord turns out to be in coincidence with the conventional quantum discord (see Supplementary Information for details).

We have studied experimentally the dynamical behavior of the quantum discord (quantum correlation), as well as the classical correlation of P:Si system, a solid-state system with genuine rather than simulated noisy environment. The results show clearly a transition from classical decoherence to quantum decoherence at a critical point $t_c \approx 166$ ns: in the initial interval $0\leq t \leq t_c$, the quantum discord does not change while the classical correlation decays, but after the critical point, the classical correlation stays constant, while the quantum discord decays. Furthermore, we have not only observed that the critical point $t_c$ can be extended via a simple DD approach by about 50 times to 8.0 $\mu$s and further to about 12 $\mu$s with multiple two-flip DD pulses, but also clearly demonstrated the decay and revival of both the classical and the quantum correlations. In view of that quantum discord could be the reason for the power of quantum computation in some cases \cite{Datta2008,Lanyon2008PRL}, a resource for remote state preparation \cite{Dakic2012natphys}, together with the demonstration of an operation method to use quantum discord as a physical resource \cite{Gu2012natphys}, here the experimental demonstration of the existence of a non-decay region, the revival and prolonging of the quantum discord in a noisy solid-state system may have great potential applications in quantum information processing.

\maketitle
\section {Supplementary information for "Experimental protection and revival of quantum correlation in open solid systems"}

\maketitle
\section {1~~STATE PREPARATION}
In order to prepare a density matrix in the form of the Bell-diagonal states which satisfy the conditions $c_1 (0)=0, |c_3 (0)| < |c_2 (0)| \ll 1$, several different microwave and radio-frequency pulses need to be designed elaborately. Starting from the thermal equilibrium state $\rho_{0}=\frac{1}{4}\mathbbm1_{4\times4}-\varepsilon \sigma_{z}\otimes \mathbbm1_{2\times2}$, where $\varepsilon = g \beta_e B_0/ {8 k_B T} = 7.35\times 10^{-3}$ (at temperature 8 K) is the ratio between the magnetic and thermal energies, a MW2 pulse is used to flip the electron spin with an angle $\theta_1$ while $I_z = 1/2$. There is a waiting time ($\tau_1 = 1~\mu$s $\gg T_{2e}^*$) followed to let the off-diagonal elements of the density matrix decay off. Then a $\theta_2$ RF1 is applied, which is followed by $\tau_2 = 200~\mu$s $\gg T_{2e}^*$. It is noted that, since both the electron and the nuclear spin longitudinal decoherence time $T_1$'s are sufficiently larger than not only the transversal docoherence time $T_2$, but also the experiment time scale, the longitudinal relaxation can be neglected, and so the diagonal elements of the density matrix remain invariant.

 \begin{widetext}
 \setlength{\mathindent}{0cm}
\begin{equation}
\label{density_matrix1}
\hspace{0mm}\rho_{1}=\frac{1}{4}\mathbbm1+\varepsilon\begin{pmatrix}
 - \sin^2\left[\frac{\theta _2}{2}\right]-\cos[\theta _1]\text{ }\cos^2\left[\frac{\theta _2}{2}\right]
& 0 & 0 & 0 \\
 0 & -\cos^2\left[\frac{\theta _2}{2}\right]-\cos[\theta _1]\sin\left[\frac{\theta _2}{2}\right]
& 0 & 0 \\
 0 & 0 & \cos[\theta _1] & 0 \\
 0 & 0 & 0 & 1
\end{pmatrix}
\end{equation}
\end{widetext}

The next $\pi/2$ pulse RF2 is to equalize the diagonal elements $\rho_{33}$ and $\rho_{44}$, along with generating some other equivalent off-diagonal elements $\rho_{34}$ and $\rho_{43}$. The coherence elements decay with a characteristic time $T_{2n}^*$.

 \begin{widetext}
 \setlength{\mathindent}{0cm}
\begin{equation}
\label{density_matrix2}
\hspace{0mm}\rho_{2}=\frac{1}{4}\mathbbm1+\varepsilon\begin{pmatrix}
- \text{sin}^2\left[\frac{\text{$\theta _2$}}{2}\right]-\text{cos}[\text{$\theta _1$}]\text{  }\text{cos}^2\left[\frac{\text{$\theta _2$}}{2}\right]
& 0 & 0 & 0 \\
 0 & -\text{cos}^2\left[\frac{\text{$\theta _2$}}{2}\right]-\text{cos}[\text{$\theta _1$}]\text{  }\text{sin}^2\left[\frac{\text{$\theta _2$}}{2}\right]
& 0 & 0 \\
 0 & 0 & \text{cos}^2\left[\frac{\text{$\theta _1$}}{2}\right] & -i \text{sin}^2\left[\frac{\text{$\theta _1$}}{2}\right] \\
 0 & 0 & i \text{sin}^2\left[\frac{\text{$\theta _1$}}{2}\right] & \text{cos}^2\left[\frac{\text{$\theta _1$}}{2}\right]
\end{pmatrix}
\end{equation}
\end{widetext}

The following $\pi/2$ pulse RF1 is to equalize the diagonal elements $\rho_{11}$ and $\rho_{22}$, along with generating some other equivalent off-diagonal elements $\rho_{12}$ and $\rho_{21}$. Since the pulse length of the radio-frequency $\pi /2$ pulse is $5~\mu$s, the decay of $\rho_{34}$ and $\rho_{43}$ in the period of pulse $\pi/2$ RF1 needs to be considered.
 \begin{widetext}
 \setlength{\mathindent}{0cm}
\begin{equation}
\label{density_matrix3}
\hspace{0mm}\rho_{3}=\frac{1}{4}\mathbbm1+\varepsilon\begin{pmatrix}
 -\text{cos}^2\left[\frac{\text{$\theta _1$}}{2}\right] & i \text{cos}[\text{$\theta _2$}] \text{sin}^2\left[\frac{\text{$\theta _1$}}{2}\right] & 0
& 0 \\
 -i \text{cos}[\text{$\theta _2$}] \text{sin}^2\left[\frac{\text{$\theta _1$}}{2}\right] & -\text{cos}^2\left[\frac{\text{$\theta _1$}}{2}\right] &
0 & 0 \\
 0 & 0 & \text{cos}^2\left[\frac{\text{$\theta _1$}}{2}\right] &-i \text{sin}^2\left[\frac{\text{$\theta _1$}}{2}\right]f(t_{\frac{\pi}{2}}) \\
 0 & 0 & i \text{sin}^2\left[\frac{\text{$\theta _1$}}{2}\right] f(t_{\frac{\pi}{2}})& \text{cos}^2\left[\frac{\text{$\theta _1$}}{2}\right]
\end{pmatrix}
\end{equation}
\end{widetext}

Where $f(t_{\frac{\pi}{2}})$ is the proportion of $\rho_{34}$ and $\rho_{43}$ decay during the $\pi/2$ pulse RF1. The last MW2 pulse is applied to transfer the off-diagonal elements $\rho_{34}$, $\rho_{43}$ to $\rho_{14}$, $\rho_{41}$, and the off-diagonal elements $\rho_{12}$, $\rho_{21}$ to $\rho_{23}$, $\rho_{32}$.
 \begin{widetext}
 \setlength{\mathindent}{0cm}
\begin{equation}
\label{density_matrix4}
\hspace{0mm}\rho_{4}=\frac{1}{4}\mathbbm1+\varepsilon\begin{pmatrix}
 \text{cos}^2\left[\frac{\text{$\theta _1$}}{2}\right] & 0 & 0 & -\text{sin}^2\left[\frac{\text{$\theta _1$}}{2}\right]f(t_{\frac{\pi}{2}}) \\
 0 & -\text{cos}^2\left[\frac{\text{$\theta _1$}}{2}\right] & \text{cos}[\text{$\theta _2$}] \text{sin}^2\left[\frac{\text{$\theta _1$}}{2}\right] &
0 \\
 0 & \text{cos}[\text{$\theta _2$}] \text{sin}^2\left[\frac{\text{$\theta _1$}}{2}\right] & -\text{cos}^2\left[\frac{\text{$\theta _1$}}{2}\right] &
0 \\
 -\text{sin}^2\left[\frac{\text{$\theta _1$}}{2}\right]f(t_{\frac{\pi}{2}}) & 0 & 0 & \text{cos}^2\left[\frac{\text{$\theta _1$}}{2}\right]
\end{pmatrix}
\end{equation}
\end{widetext}

A suitable choice of the angles $\theta _1$ and $\theta _2$ satisfying the conditions $\text{cos}[\text{$\theta _2$}]  =f(t_{\frac{\pi}{2}})$ and $\text{cos}^2\left[\frac{\text{$\theta _1$}}{2}\right]<\text{cos}[\text{$\theta _2$}] \text{sin}^2\left[\frac{\text{$\theta _1$}}{2}\right]$ generates the final  density matrix in the form of the Bell-diagonal state, which satisfies the conditions $c_1 (0)=0, |c_3 (0)| < |c_2 (0)| \ll 1$. In our experiment, the actual values of $\theta _1$ and $\theta _2$ are chosen to be $\theta _1\thickapprox0.70~\pi$ and $\theta _2\thickapprox0.28~\pi$. So the theoretical density matrix after the five pulses becomes
 \begin{widetext}
 \setlength{\mathindent}{0cm}
\begin{equation}
\label{density_matrix5}
\hspace{0mm}\rho_{5}=\frac{1}{4}\mathbbm1+\varepsilon\begin{pmatrix}
 0.206 & 0 & 0 &- 0.506 \\
 0 &- 0.206 &0.506 &
0 \\
 0 & 0.506 &- 0.206 &
0 \\
-0.506 & 0 & 0 &0.206
\end{pmatrix}
\end{equation}
\end{widetext}

\maketitle
\section {2~~STATE TOMOGRAPHY AND THE DECAY OF OFF-DIAGONAL ELEMENTS}

To identify the dynamics of classical and quantum correlations, besides the state preparation, the state tomography is needed throughout the state evolution. The state tomography of two qubit in solid is used and introduced in detail in\cite{Rong2012PRB}. Here in our experiment, the density matrix of is measured after  the state preparation to be
 \begin{widetext}
 \setlength{\mathindent}{0cm}
\begin{equation}
\label{density_matrix exp}
\hspace{0mm}\rho_{AB}=\frac{\mathbbm1}{4}+\varepsilon \begin{pmatrix}
 0.21(3) & 0.01(1)+i0.02(1) & -0.01(1)-i0.02(3) & -0.54(8)+i0.11(1) \\
0.01(1)-i0.02(1) & -0.21(3) & 0.50(8)-i0.15(2) & 0.02(1)+i0.01(1) \\
-0.01(1)+i0.02(3) & 0.50(8)+i0.15(2) & -0.17(3) & 0.01(2)-i0.02(2) \\
-0.54(8)-i0.11(1) & 0.02(1)-i0.01(1) & 0.01(2)+i0.02(2) & 0.17(3)
\end{pmatrix}
\end{equation}
\end{widetext}
The quoted errors come from the fitting errors for the nuclear nutation.

\begin{figure}
\centering
\includegraphics[width=0.8\columnwidth]{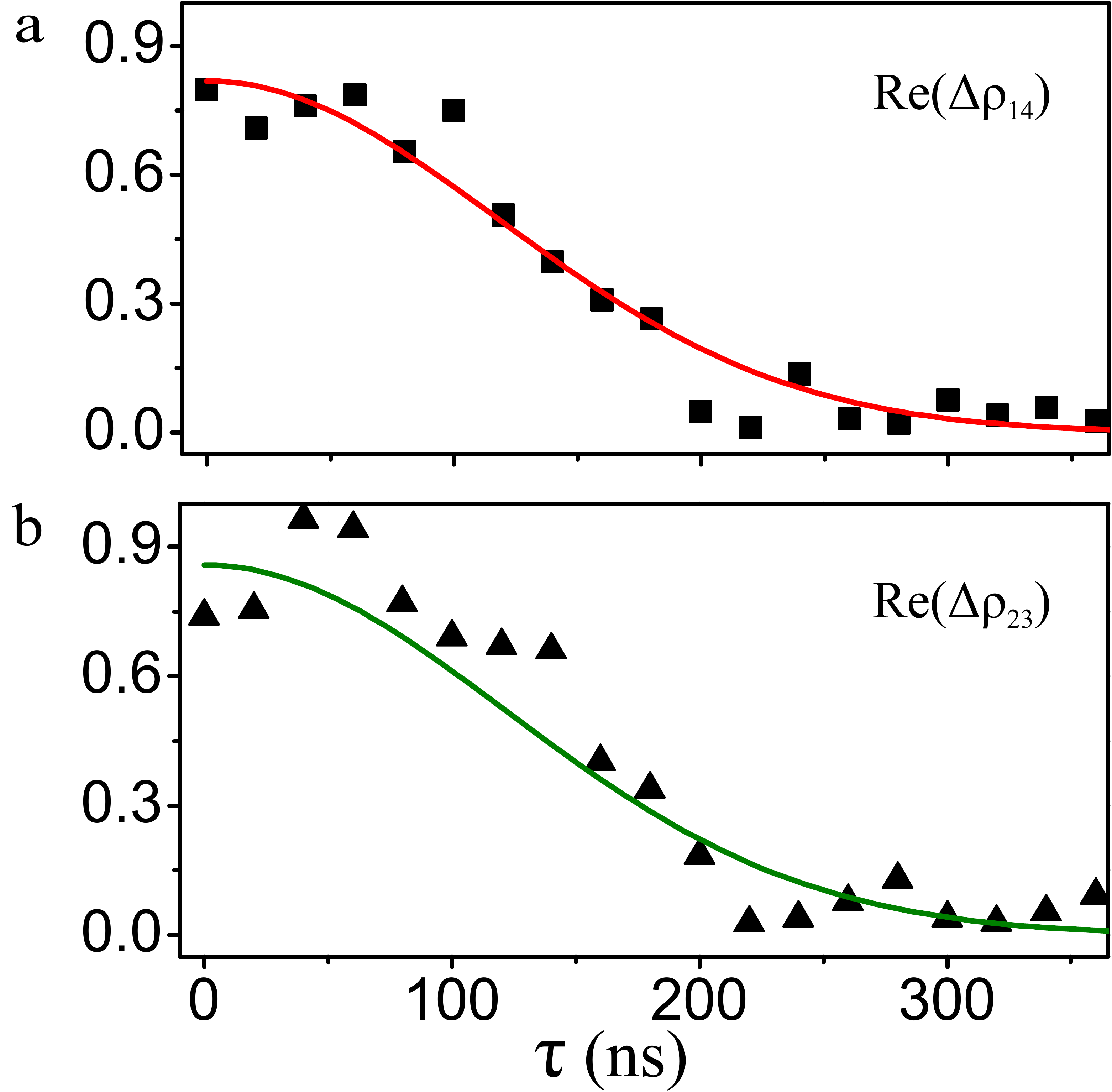}\caption{\label{figs1} {\bf The decay of the real part of $\rho_{14}$ and $\rho_{23}$.} {\bf a, b,} The decay of the real part of $\rho_{14}$ and $\rho_{23}$, in unit of $\varepsilon$, respectively. The solid line is the fitted function of $\sim \exp[-(t/T_{\rm decay})^2]$, with $T_{\rm decay}=175$ ns.
}
\end{figure}

According to Eq.(4) (main text) and Eq.(5) (main text), the discord and classical correlation of the states which we are studying just depend on the decay of $\rho_{23}$, $\rho_{14}$ and their relationship with the diagonal elements. The decay of $\rho_{23}$ and $\rho_{14}$ are measured as Fig.\ref{figs1}a,b shows, in unit of $\varepsilon$, respectively. The solid line is the fitted function of $\sim \exp[-(t/T_{\rm decay})^2]$, with $T_{\rm decay}=175$ ns.

\maketitle
\section {3~~CALCULATION OF CORRELATIONS}

Here we take the reconstructed density matrix of Eq.(\ref{density_matrix exp}) as an example, i.e. the one shown in Fig. 1c,d (main text).
The mutual information of $\rho_{AB}$ is  calculated with $I(\rho_{AB})=S(\rho_A)+S(\rho_B)-S(\rho_{AB})$ to be $2.0(6)\times10^{-4}$. The classical correlation and quantum discord are obtained with $C_B(\rho_{AB})=\max_{ \{\Pi_k  ^B\}}[S(\rho_A)-\sum_k p_k S(\rho^{k}_A)]$ and $D_B(\rho_{AB})=I(\rho_{AB})-C_B(\rho_{AB})$ by optimizing over all 1-qubit measurements. Since it has been proven that the optimal measurement is always projective for two-qubit states, it is enough to maximize over all the following projective measurements $\{\mathbbm1 \otimes \mid\Theta_k><\Theta_k\mid,k=\parallel,\perp\}$, where $\mid\Theta_{\parallel}\rangle =\cos\theta\mid0\rangle +e^{i\phi}\sin\theta|1\rangle $ and $|\Theta_{\perp}\rangle =e^{-i\phi}\sin\theta|0\rangle-\cos\theta|1\rangle$ presents an arbitrary basis of B formed by two orthogonal states on the Bloch sphere, with $0\leq\theta\leq\frac{\pi}{2}$ and $0\leq\phi\leq2\pi$. The classical correlation and quantum discord are calculated numerically to be $1.8(6)\times10^{-4}$ and $2(1)\times10^{-5}$. The errors of the correlations are obtained with the maximum difference of correlations between density matrices $\rho_{AB}^{\prime}$ and $\rho_{AB}$, where each element of $\rho_{AB}^{\prime}$ falls near that of $\rho_{AB}$ within the element errors.

\maketitle
\section {4~~DYNAMICS OF GEOMETRIC MEASURE OF QUANRUM DISCORD}

There are many ways to quantify and verify quantum correlations \cite{Modi2012arXiv}, such as quantum deficit, distillable common randomness, measurement-induced disturbance, symmetric discord, relative entropy of discord and dissonance and geometric measures. Recently, it has been shown that the
geometric measure of quantum discord is related to the fidelity of quantum remote state preparation, which provides its operational meaning \cite{Dakic2012natphys}. Herein we focus the dynamics of geometric measure of quantum discord which first introduced by Daki$\acute{c}$, Vedral, and Brukner \cite{Dakic2010PRL}. This measure is significant in capturing quantum correlations from a geometric perspective and can be evaluated explicitly and leads to an explicit formula for any two-qubit state. It is defined as the normalized trace distance to the set of classical states\cite{Dakic2012natphys,Dakic2010PRL}
\begin{equation}
\label{GQD}
\hspace{0mm} \mathcal{D}^2(\rho_{AB})=2~\text{min}_{\chi \in \Omega_{0}}\| \rho_{AB}-\chi \|^2=2~\text{min}_{\chi \in \Omega_{0}}\text{Tr}(\rho_{AB}-\chi)^2,
\end{equation}
where $\Omega_{0}$ denotes the set of zero-discord states.

For evolutive Bell-diagonal states Eq.(2) in the main text, the geometric measure of quantum discord can be expressed as

\begin{equation}
\label{GQDe}
\mathcal{D}^2[\rho(t)]=\frac{1}{2}~(c^2_1(t)+c^2_2(t)+c^2_3(t)-\text{max}\{c^2_1(t),c^2_2(t),c^2_3(t)\}),
\end{equation}
In the case of $c_1 (0)=0, |c_3 (0)| < |c_2 (0)| \ll 1$, it is calculated to be

\begin{equation}
 \label{GQDt}
\mathcal{D}^2[\rho(t)]=
   \begin{cases}
  \frac{1}{2}~c^2_3 & \text{if $t\leqslant t_c$},\\
  \frac{1}{2}~c^2_2 (t) & \text{if $t>t_c$}.
  \end{cases}
\end{equation}
Where $t_c$ is same to the main text $t_c= \sqrt {-\ln[c_3/c_2(0)]} T_{2e}^\ast$. Comparing Eq.(\ref{GQDt}) with Eq.(5) (main text), we see that the difference between geometric measure of quantum discord and quantum discord is just a constant coefficient 1/ln2. Consequently, in the noisy environment, geometric measure of quantum discord will be constant in the initial period $t\leqslant t_c$ , while after the period $t_c$, it will be reduced sharply.

\begin{figure}
\centering
\includegraphics[width=0.9\columnwidth]{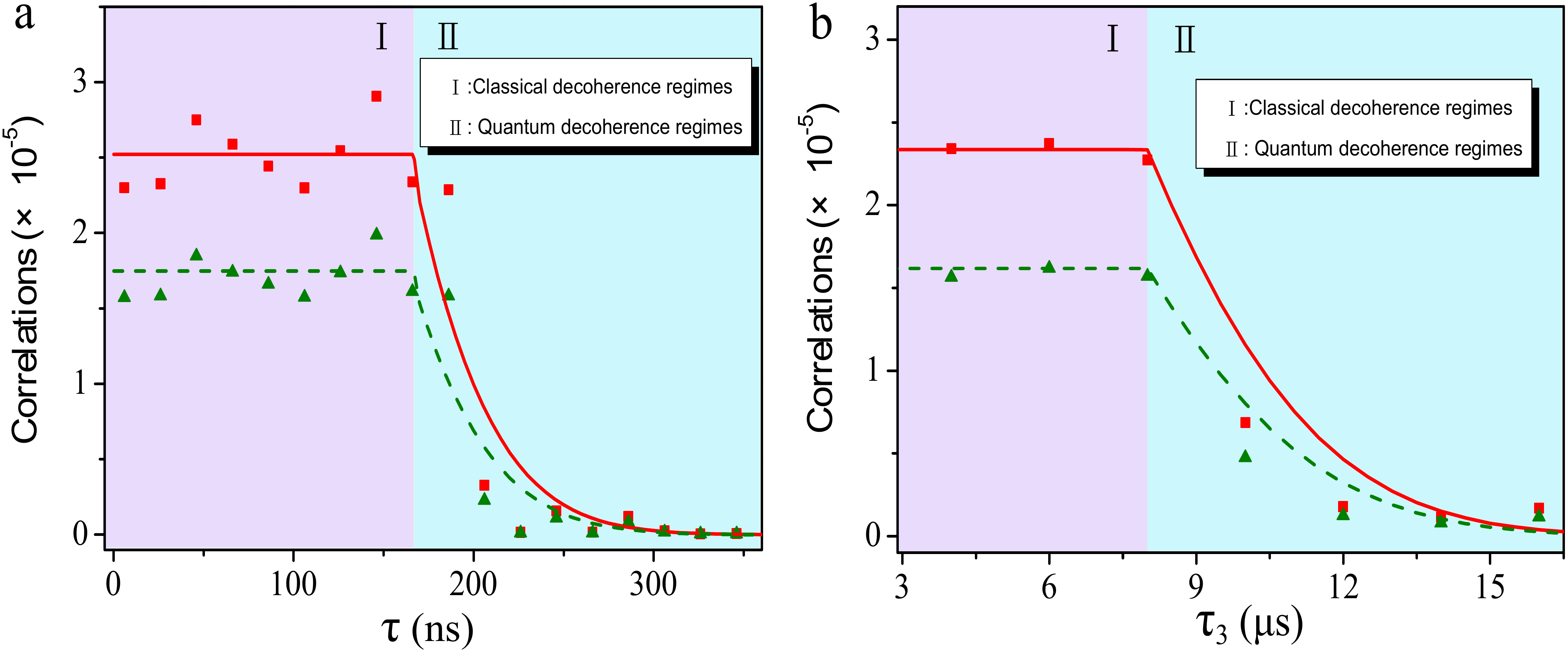}\caption{\label{figs2}  {\bf  Dynamics of quantum discord and geometric measure of quantum discord.} {\bf a,}  Quantum discord (red square) and geometric measure of quantum discord (darkgreen   triangle) are numerical computed with their original definitions. The curves are the theoretical predication according to Eq.(5) (main text) and Eq.(\ref{GQDt}). \uppercase\expandafter{\romannumeral1} and \uppercase\expandafter{\romannumeral2} stand for the constant and decoherence regimes, respectively.  {\bf b,} Preservation of quantum discord and geometric measure of quantum discord by dynamics decoupling. The curves are drawn to follow the trend of the variation of correlations.
}
\end{figure}

 Fig.\ref{figs2}a shows the dynamics of quantum discord and geometric measure of quantum discord. The values of Quantum discord (red square) and geometric measure of quantum discord (darkgreen   triangle) are numerical computed with their original definitions.  There is sudden transition of quantum discord from constant to decoherence regimes occurs at about 166 ns as well as geometric measure of quantum discord. In the initial period ($t\lesssim 166$ ns), both the quantum discord and geometric measure of quantum discord remain constant, while for $t\gtrsim166$ ns, they decrease dramatically. Fig.\ref{figs2} shows that the decay of both quantum discord and geometric measure of quantum discord become slower, and the period before the sudden transition of correlations is prolonged by DD from about 166~ns to $8~\mu$s.

\end{document}